\documentclass{PoS}

\usepackage{graphicx} 
\input babarsym

\newcommand{\text}[1]{\mbox{\rm #1}}
\newcommand{\gfrac}[2]{\displaystyle\frac{#1}{#2}}
\newcommand{\dd}{\mbox{d}}

\title{Measurement of $e^+e^-$ \to hadrons cross sections at
 BABAR, and implication for the muon $g-2$}

\ShortTitle{$e^+e^-$ \to hadrons cross sections at BABAR and the muon
 $g-2$}

\author{\speaker{Denis Bernard}\thanks{On behalf of the \babar\ Collaboration}\\
Laboratoire Leprince-Ringuet, Ecole Polytechnique, CNRS/IN2P3, F-91128 Palaiseau, France \\
 E-mail: \email{denis.bernard@in2p3.fr}}

\abstract{The \babar\ Collaboration has an intensive program of
 studying hadronic cross sections at low-energy \epem collisions,
 accessible via initial-state radiation. Our measurements
 allow significant improvements in the precision of the predicted
 value of the muon anomalous magnetic moment, that
 shed light on the current $\approx$ 3.5 sigma difference
 between the predicted and the experimental values. We have published
 results on a number of processes with two to six hadrons in the
 final state. We report here the results of recent studies with the
 final states that constitute the main contribution to the hadronic
 cross section below 3 GeV, as 
 $\epem \to \pip\pim$, $\Kp\Km$, $\KS\KL$ and $\epem \to 4$ hadrons.
}

\FullConference{XV International Conference on Hadron Spectroscopy-Hadron 2013\\
		4-8 November 2013\\
		Nara, Japan }

\begin{document}

The last decades have seen an increase of the precision of both the
measurement and of the theoretical understanding of the magnetic
moment of the muon $g_\mu$, presently one of the most precisely known
quantity in physics.
The ``anomalous'' magnetic moment, i.e. the deviation 
$a_\mu \equiv (g_\mu -2)/2$ of $g_\mu$ from the value of $g=2$ for a
pointlike Dirac particle is presently measured to
$a_\mu(\expt) = (11 659 208.0 \pm 5.4 (\stat) \pm 3.3 (\syst)) \times 10^{-10}$
\cite{Bennett:2006fi}, while the prediction is 
close to 
$a_\mu(\theo) = (11 659 181 \pm 5 (\stat) \pm 1 (\syst)) \times 10^{-10}$, 
where I dared compute an average of the predictions from
various authors,
\cite{Davier:2010nc,Hagiwara:2011af,Jegerlehner:2011ti}, 
and I give as statistical uncertainty the typical uncertainty of
each prediction and as systematics uncertainty an estimate of the
variation amongst these three references.
The measured value exceeds the prediction by 
$\Delta a_\mu = (27 \pm 8) \times 10^{-10}$, 
assuming Gaussian statistics, that is an
$\approx 3.3$ standard deviation effect.

The computation of $a_\mu$ involves a perturbative development and the
QED contributions is obviously the main contribution to $a_\mu$, but
its precision is actually extremely small, equal to 
$8 \times 10^{-13}$ from the recent 10th order (in $e$) calculation of Ref.
\cite{Aoyama:2012wk}.
The uncertainty of the present prediction of $a_\mu$ is actually dominated
by the contribution of the hadronic vacuum polarization (VP).
As is well known, QCD is not suited to low energy calculations. 
Therefore the VP contribution to $a_\mu$ is computed from the
``dispersion integral'' 
(see the detailed presentation at \cite{Jegerlehner:2009ry}) : 
\begin{eqnarray}
a_\mu^{\had} = \left( \gfrac{\alpha m_\mu} {3\pi} \right)^2
\int{\gfrac{R_{\had}(s) \times \hat{K}(s)}{s^2} \dd s}, 
\end{eqnarray}
where $ R_{\had}(s) $ is the the cross section of $\epem$ to hadrons
at center-of-mass (CMS) energy squared $s$, normalized to the
pointlike cross section :
$ R_{\had}(s) = \sigma_{\epem\to hadrons} / \sigma_{\epem\to \mumu}$,
and $\hat{K}(s)$ is a known function that is of order unity on the $s$ range 
$[(2 m_\pi c^2)^2, \infty]$.
Technically, the lowest energy part of the integral is obtained from
experimental data (currently up to $E_{cut} = 1.8 \gev$), while the
high-energy part is computed from perturbative QCD.
Due to the presence of the $s^2$ factor at the denominator of the
integrand, the precision of the prediction of $a_\mu$ relies on
precise measurements at the lowest energies, and the channels
 $\pip \pim$, 
 $\pip \pim \piz$, 
 $\pip \pim 2\piz$, $\pip \pim\pip \pim$, 
 $K K$ are of particular importance.
\begin{figure}
\centerline{
\includegraphics[width=0.76\linewidth]{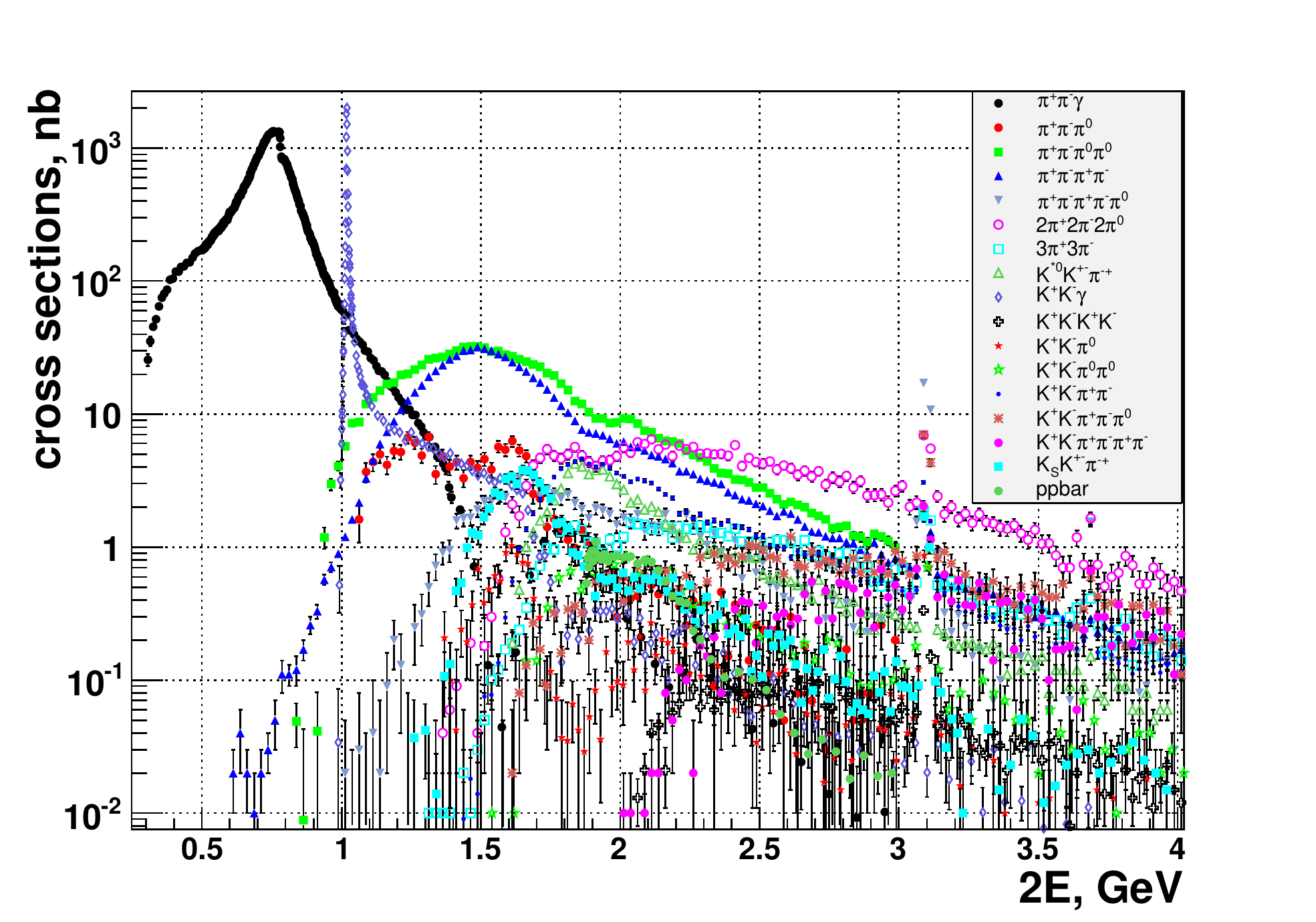}
}
\caption{Summary of the \babar\ measurements 
(The $\pip\pim\piz\piz$ entry is preliminary \cite{Druzhinin:2007cs}).
Thanks to Fedor V. Ignatov.
\label{fig:summary:oct2013}
}
\end{figure}

\begin{table}
\footnotesize
\begin{center} \footnotesize
\begin{tabular}{llllllllll}
\hline 
 \hline \noalign{\vskip3pt}
Channels & $\cal L$ \invfb & method & reference \\
\hline 
 \hline \noalign{\vskip3pt}
$\KS\KL$, {\bf 
$\KS\KL \pip\pim$, 
$\KS\KS \pip\pim$, 
$\KS\KS \Kp\Km $} & 469 & LO & preliminary
\\
$\antiproton p $ & 454 & LO & \cite{:2013xe} 
\\
$\antiproton p $ & 469 & no tag LO & \cite{Lees:2013uta} 
\\
$\Kp \Km$ & 232 & NLO & \cite{Lees:2013gzt} 
\\
$2(\pip\pim)$ & 454 & LO & \cite{Lees:2012cr} 
\\
$\Kp\Km \pip\pim$, 
$\Kp\Km \piz\piz$, 
$\Kp\Km \Kp\Km $ & 454 & LO & \cite{Lees:2011zi} 
\\
$\pip\pim$ & 232 & NLO & \cite{Aubert:2009ad} \cite{Lees:2012cj} 
\\
$\Kp \Km \eta$, 
$\Kp \Km \piz$, 
$\Kz \Kpm \pimp$ & 232 & LO & \cite{Aubert:2007ym} 
\\
 $\pip\pim\piz\piz$ & 232 & LO & \cite{Druzhinin:2007cs} preliminary 
\\
$2(\pip\pim)\piz, $
{\bf 
$2(\pip\pim)\eta$},
{\bf $\Kp \Km \pip \pim \piz$},
 {\bf 
$\Kp \Km \pip \pim \eta$} & 232 & LO & \cite{Aubert:2007ef} 
\\
{\bf 
$\Lambda \overline \Lambda $,
$\Lambda \Sigma^0$,
$ \Sigma^0 \Sigma^0$
}
& 232 & LO & 
\cite{Aubert:2007uf} 
\\
$3(\pip\pim)$, 
$2(\pip \pim \piz)$, {\bf 
$\Kp \Km 2(\pip\pim)$} & 232 & LO & \cite{Aubert:2006jq} 
 \\
$ \pip \pim \piz $ & 89 & LO & \cite{Aubert:2004kj} 
\end{tabular}
\end{center}

\caption{Summary of the \babar\ results on ISR production of exclusive
 hadronic final states
(publications that have been superseded by updated results with a
larger dataset are removed).
\label{tab:compilation}
}
\end{table}

The \babar\ experiment has committed itself to the systematic
measurement of the production of all hadronic final states within
reach in the relevant energy range ($E < E_{cut}$) over the last
decade, using the initial state radiation (ISR) process 
(Fig. \ref{fig:summary:oct2013}, Tab. \ref{fig:summary:oct2013}).
The cross section of the $\epem$ production of a final state $f$ at a
CMS energy squared $s'$ can be obtained from the differential
cross section of the ISR production $\epem \to f \gamma$ through the
expression :
\begin{eqnarray}
\gfrac{\dd \sigma_{[\epem \to f \g]}}{\dd s'} (s')
=
\gfrac{2m}{s} W(s, x) \sigma_{[\epem \to f]} (s'),
\end{eqnarray}
where $W(s, x)$, the density of probability to radiate a photon with
energy $E_\g = x\sqrt{s}$ is a known ``radiator'' function
\cite{Binner:1999bt}, and $s$ is here the CMS energy squared of the initial
\epem pair.
In contrast with the energy scans that provided the earlier
experimental information on the variations of $R$, this ISR method
allows a consistent measurement on the full energy range with the same
accelerator and detector conditions.
The observation of the hadronic final state alone would allow the
reconstruction of the event and the measurement of $s'$, but in
addition the observation of the ISR photon ($\gamma$-tagging) provides
a powerful background rejection and a good signal purity.

In the case of \babar\, the \epem initial state is strongly boosted so
that the reconstruction efficiency is large down to threshold.
Most of these measurements have used a leading-order (LO) method, in
which the final state $f$ and the ISR photon are reconstructed
regardless of the eventual presence of additional photons.
The differential luminosity is obtained from the luminosity of the
collider, known with a typical precision of $1 \%$, and involves a
computation of the detection efficiency that relies on Monte Carlo
(MC) simulations
 \cite{Aubert:2004kj,Aubert:2006jq,Aubert:2007uf,Aubert:2007ef,Aubert:2007ym}.
This experimental campaign has lead \babar\ to improve the precision
of the contribution to $a_\mu$ of most of the relevant channels by a
large factor, typically close to a factor of three.
More recently \babar\ has developped a new method that was applied to
the dominant channel $\pip\pim$ \cite{Aubert:2009ad,Lees:2012cj}
and to the $\Kp\Km$ channel \cite{Lees:2013gzt}.
The control of the systematics below the \% level made it necessary to
perform the analysis at the NLO level, that is to take into account
the possible radiation of an additional photon
 ($\epem \to f \gamma_{ISR} (\gamma))$.
The impossibility to control the global differential luminosity with
the desired precision, in particular the MC-based efficiency, lead us
to derive the value of $R$ from the ratio of the ISR production of the
final state $f$ to the ISR production of a pair of muons,
$\mu^+\mu^-$.
Most of the systematics, including those related to the absolute
luminosity, of the ISR photon reconstruction, of additional ISR
radiation, cancel in the ratio.
 \babar\ also performed updates of former
works, using the full data set for the 
$\pip\pim\pip\pim$ \cite{Lees:2012cr},
$p \bar{p}$ \cite{:2013xe}, 
$\Kp\Km\pip\pim$, $\Kp\Km\piz\piz$ and $\Kp\Km\Kp\Km$ \cite{Lees:2011zi}
channels, 
and channels with two neutral kaons 
($\KS\KL$, 
 $\KS\KL \pip\pim$,
 $\KS\KS \pip\pim$,
 $\KS\KS \Kp\Km$)
\cite{ref:prel}
using the LO method.
The $p \bar{p}$ measurement has also been extended up to 6.5 \gev by an
untagged analysis \cite{Lees:2013uta}.
\begin{table} 
\footnotesize
\begin{tabular}{l|lllllllll} 
\hline
\hline
 & $\pip\pim$ & $\pip\pim\pip\pim$ & $\Kp\Km$ 
\\
\hline
\hline
\babar\ & $514.1 \pm 2.2 \pm 3.1$ \cite{Aubert:2009ad} \cite{Lees:2012cj} & $22.93 \pm 0.18 \pm 0.22 \pm 0.03 $ \cite{Lees:2012cr} & $13.64 \pm 0.03 \pm 0.36$ \cite{Lees:2013gzt} 
\\
Previous average
\cite{Davier:2010nc} & $503.5 \pm 4.5$ & $21.63 \pm 0.27 \pm 0.68$ & $13.35 \pm 0.10 \pm 0.43 \pm 0.29$
\\
Their difference $\Delta$ & $+10.6 \pm 5.9$ & $+1.30 \pm 0.79$ & $+0.29 \pm 0.63$
\end{tabular}
\caption{Contributions to $a_\mu$ for recent \babar\ publications : comparison of the measured value to the previous world average on the energy range $< 1.8 \gev$ (units $10^{-10}$).
\label{tab:comparison}
}
\end{table}
The \babar\ results for the contributions to $a_\mu$ for 
$\pip\pim$, $\pip\pim\pip\pim$ and $\Kp\Km$ 
are larger than, 
have a similar to or better precision than, 
and are compatible within less than two standard deviation with, 
the world combination of previous results (Table \ref{tab:comparison}).

\begin{figure}
\centerline{
\includegraphics[width=0.9\linewidth]{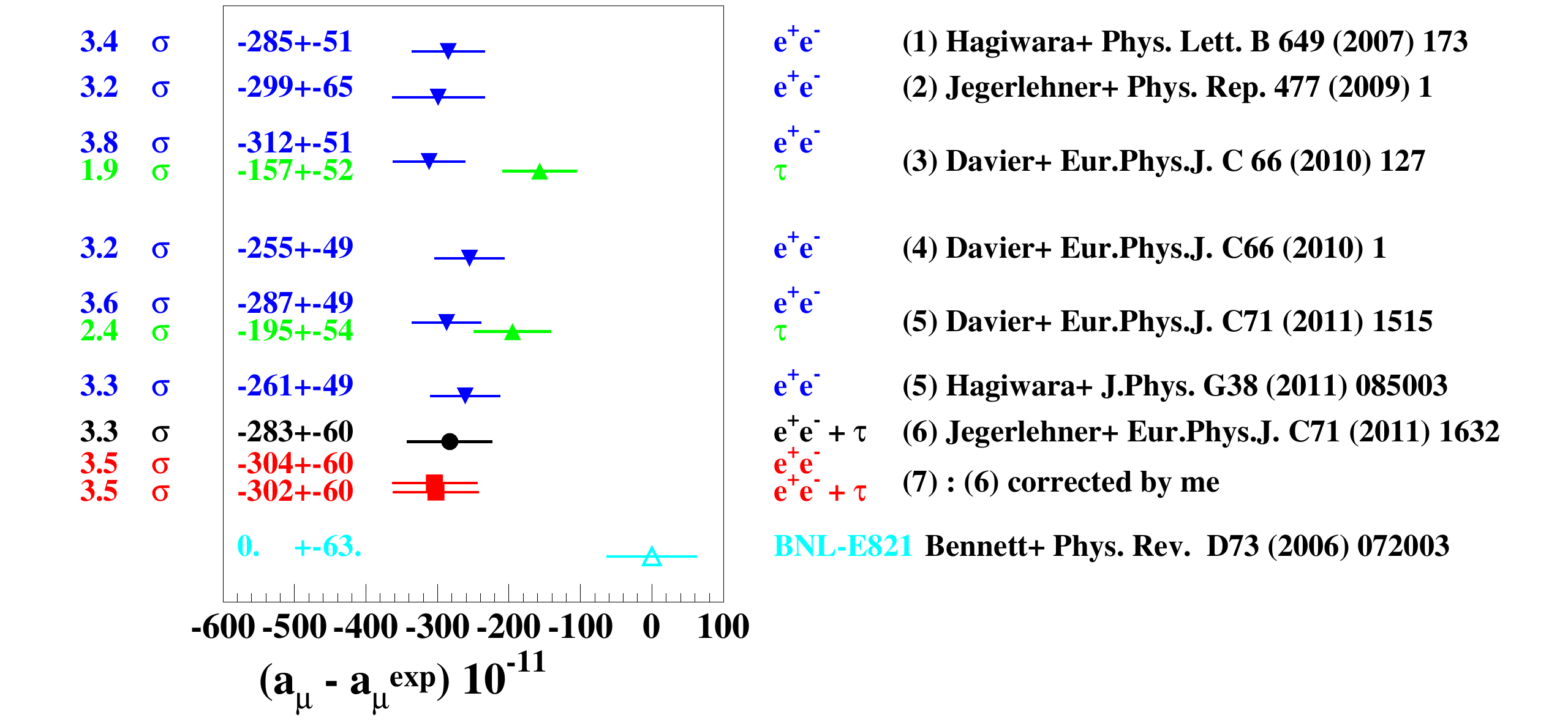}
}
\caption{
Recent predictions of the value of $a_\mu$ in chronological order, after the experimental value is subtracted.
\Blue Blue \Black: \epem-based; 
\Green Green \Black : $\tau$ spectral function-based; 
 Black : \epem and $\tau$ combinations;
\Red Red \Black : ``corrected'' by me, see text.
\label{fig:combination}
}
\end{figure}

It is interesting to compare the evolution of the prediction of
$a_\mu$ with the availability of experimental results of increasing
precision and with the development of combination techniques.
In Fig. \ref{fig:combination} I show the value of the most recent
predictions, after subtraction of the experimental value (units
$10^{-11}$).
The most recent works of the three main groups
\cite{Davier:2010nc,Hagiwara:2011af,Jegerlehner:2011ti}
make use of our results up to and including $\pip\pim$ 
 \cite{Aubert:2009ad,Lees:2012cj}, but 
are not yet using our recent 
\cite{:2013xe,Lees:2013gzt,Lees:2012cr,Lees:2011zi}.
Given the absence of contribution of the 
$\antiproton p $ \cite{:2013xe} and 
$\Kp\Km \Kp\Km $ \cite{Lees:2011zi} channel below 1.8 \gev, 
and given the smallness of the difference $\Delta$ for the 
$\Kp \Km$ \cite{Lees:2013gzt} 
and
$2(\pip\pim)$ \cite{Lees:2012cr} 
channels, we see that including our recent results 
\cite{:2013xe,Lees:2013gzt,Lees:2012cr,Lees:2011zi} will barely change the
prediction central value.
It is reassuring to note that :
\begin{itemize}
\item
 these three predictions performed independently to a large extent, as
 far as the VP is concerned, provide results compatible with each
 other \footnote{In (7) of Fig. \ref{tab:compilation} I dared 
 correct the prediction (6) from \cite{Jegerlehner:2011ti} to align
 the $\mu_\mu / \mu_p$ value and the calculation of light-by-light
 scattering on that of the two other groups \cite{Davier:2010nc} and
 \cite{Hagiwara:2011af}.}
within a couple of $10^{-10}$;
\item
 after $\rho-\gamma$ mixing is taken into account, the discrepancy
 between the combinations based on \epem results and those based on
 the $\tau$ decay spectral functions (see Ref. \cite{Davier:2009ag})
 is solved \cite{Jegerlehner:2011ti}.
\end{itemize} 
The discrepancy between the prediction and the measurement still sits
close to 3.3 -- 3.6 standard deviations.
Given that the precision of most measurements is now dominated by the
systematics contribution there is most likely no hope of major
improvements from a super $B$ factory like Belle 2.
Indeed, new measurements of $a_\mu$ at Fermilab \cite{Venanzoni:2012qa}
and at
J-PARC \cite{Mibe:2011zz} are eagerly awaited.


\end{document}